\documentclass{elsart}

 \usepackage{graphicx}

\usepackage{amssymb}

\journal{Physica A}

\begin{document}

\begin{frontmatter}



\title{Effective field theory for the Ising model with a fluctuating exchange
integral in an asymmetric bimodal random magnetic field: A differential
operator technique}

\author{Ioannis A. Hadjiagapiou\corauthref{cor1}}
\corauth[cor1]{Corresponding author.}
\ead{ihatziag@phys.uoa.gr}

\address{Section of Solid State Physics, Department of Physics,
University of Athens, Panepistimiopolis, GR 15784 Zografos,
Athens, Greece}

\begin{abstract}

The spin-1/2 Ising model on a square lattice, with fluctuating bond
interactions between nearest neighbors and in the presence of a random
magnetic field, is investigated within the framework of the effective field
theory based on the use of the differential operator relation. The random
field is drawn from the asymmetric and anisotropic bimodal probability distribution
$P(h_{i})=p \delta(h_{i}-h_{1}) + q \delta (h_{i}+ ch_{1})$, where the site
probabilities $p,q$ take on values within the interval $[0,1]$ with the
constraint $p+q=1$; $h_{i}$ is the random field variable with strength
$h_{1}$ and $c$ the competition parameter, which is the
ratio of the strength of the random magnetic field in the two
principal directions $+z$ and $-z$; $c$ is considered to be positive
resulting in competing random fields. The fluctuating bond is drawn from the
symmetric but anisotropic bimodal probability distribution
$P(J_{ij})=\frac{1}{2}\{\delta(J_{ij}-(J+\Delta)) + \delta
(J_{ij}-(J-\Delta))\}$, where $J$ and $\Delta$ represent the average
value and standard deviation of $J_{ij}$, respectively.
We estimate the transition temperatures, phase diagrams (for various
values of system's parameters $c,p,h_{1},\Delta$), susceptibility,
equilibrium equation for magnetization, which is solved in order to
determine the magnetization profile with respect to $T$ and $h_{1}$.

\end{abstract}
\date{\today}

\end{frontmatter}

\newpage
\section{Introduction}

Prediction of the critical behavior of modified spin models in
the presence of site or bond dilution, random bonds, random fields
has been the subject of many studies in the last decades;
this modification brings about considerable changes in the critical
behavior of these systems, such as replacement of a first-order
phase transition (FOPT) by a second-order phase transition (SOPT),
depression of tricritical points and critical end points, new
critical points and universality classes, etc \cite{rsdim,rbim,huiberker}.
The study of the aforementioned disordered systems is based on the
standard models, such as Ising, Blume-Capel, Baxter-Wu, Heisenberg,
etc, modified accordingly to meet the requirements under
consideration. Furthermore, extensions and versions of these
models can be applied to describe many other different situations,
such as multicomponent fluids, ternary alloys, $^{3}$He -$^{4}$He
mixtures, in addition to the magnetic systems for which these were
initially conceived. The most extensively investigated model in
statistical and condensed matter physics is the spin-1/2 Ising
model (IM), since its two dimensional version, without an external
magnetic field, was analytically solved by Onsager; it is a
simple one relative to other models of cooperative phenomena and
has a wide range of applicability to real physical systems.
Its three-dimensional version has not yet been solved
exactly, for which the only existing results are either from the
renormalization group calculations or series expansions and are
thus considered to be the "exact" ones. In its modified
versions, it exhibits a variety of multicritical phenomena, such
as a phase diagram with ordered ferromagnetic and disordered
paramagnetic phases separated by a transition line that changes
from an SOPT to an FOPT joined at a tricritical point (TCP);
besides these, critical points, critical end points, ordered
critical points of various orders, re-entrance can appear as in
the presence of random fields. The multicritical phenomena appear
in systems presenting competition among distinct types of ordering
and there are numerous circumstances in which this kind of
phenomenon can arise. In ferromagnetic systems in the presence of
random fields, the competition is between the parallel and random
ordering, causing, occasionally, the conversion of a continuous
transition into an FOPT and the subsequent appearance of TCP as
well as re-entrance in some cases \cite{asym2}.
Random field effects on magnetic systems have been systematically
studied not only for their own theoretical study but for their
experimental importance, as well.

The methods used for the study of the IM are the mean field
approximation (MFA), high temperature expansions, series
expansion, renormalization group, Monte Carlo, effective field
theory (EFT). The eldest one is the MFA, which is very popular
because of its simplicity and has played an important role
for the description of cooperative phenomena for many years; it
gives qualitative agreement with experimental results for many of
the physical quantities involved in a phase transition. However,
MFA has some unsatisfactory results in describing correctly the critical
region because of the omission of correlations when its results
are compared with those of experiments. MFA is still implemented
because it is not complicated and can give a general view of
the expected behavior for the system under consideration. Many
efforts were attempted towards improving MFA, such as the effective
field theory by Honmura and Kaneyoshi \cite{honkan}. EFT relies on
introducing a differential operator into the exact spin correlation
function identities obtained by Callen \cite{callen} and using the
van der Waerden spin identities, which improves substantially on the
standard MFA \cite{satsa,satsahon}. The EFT approach is general and may be applied
to systems with any spin value, adapting the van der Waerden identities
accordingly. This procedure presents a great versatility and has been
applied to several occasions such as pure, site- and bond-random Ising
model, although this procedure shall not yield accurate values for
the physical quantities in the critical region due to the absence
of long range fluctuations. EFT has already been used in numerous physical
problems as a tool to study the magnetic behavior of complex spin systems, such as
diluted ferromagnets \cite{kanfit,kantamhon,tagg}, pure anisotropic systems \cite{kantam},
disordered systems \cite{zheng,yusuf}, cylindrical nanowires \cite{kan}.

The Hamiltonian we shall adopt is that of Ising model with nearest
neighbor interactions, in which the bond $J_{ij}$ between two
neighboring spins varies from pair to pair randomly; the system is also under
the influence of a random field $h_{i}$, varying from site to site;
both random variables are drawn from a suitable probability distribution
function (PDF).

The Hamiltonian describing the above system is

\begin{equation}
 H=-\sum_{<i,j>}J_{ij}S_{i}S_{j}-\sum_{i}h_{i}S_{i}  \hspace{2mm},
    \hspace{20mm} S_{i}=\pm1 \label{rham}
\end{equation}

\noindent The summation in the first term extends over all nearest
neighbors and is denoted by $<i,j>$; in the second term $h_{i}$
represents a random field that couples to the one dimensional spin
variable $S_{i}$. The Hamiltonian describes the competition
between the long-range order (expressed by the first summation)
and the random ordering fields. The presence of random fields
requires two averaging procedures, the usual thermal average,
denoted by angular brackets $\langle...\rangle$ and disorder
average over the random fields denoted by $\langle...\rangle_{r}$
for the respective PDF's, which are usually a version of the
bimodal or Gaussian distributions.

The random exchange integral $J_{ij}$ between the sites $i$ and
$j$ is drawn from the symmetric and anisotropic bimodal PDF,

\begin{equation}
P(J_{ij})=\frac{1}{2}\{\delta(J_{ij}-(J+\Delta)) + \delta
(J_{ij}-(J-\Delta))\}  \label{bimodalJ}
\end{equation}

where $J$ and $\Delta$ represent the average value and standard
deviation of $J_{ij}$, respectively, implying that the exchange
integral is fluctuating. This choice of the PDF
implies that both values ($J-\Delta, J+\Delta$) of the random
exchange integral are equally probable.

The PDF for the random fields $h_{i}$ is also of the bimodal type

\begin{equation}
 P(h_{i})=p\delta(h_{i}-h_{1}) + q \delta (h_{i}+h_{2})  \label{bimodal}
\end{equation}

where $h_{2}=ch_{1}$, $c$ is the competition parameter
and is considered to be positive so that the random fields
are competitive. The site probability $p$ is the fraction of
lattice sites having a random magnetic field with strength $h_{1}$,
while the rest sites have a field with strength $(-h_{2})$ with site
probability$q$ such that $p+q=1$ and the usual choice was $p = q = \frac{1}{2}$,
symmetric case, \cite{asym2,aharony,andelman1,kaufkan}.
As far as the PDF (\ref{bimodal}) is non symmetric, the mean value
for $h_{i}$ does not vanish a priori, but, instead, is
$ <h_{i}>_{h} =  (p-cq)h_{1}$; an immediate result is that the
system is under the influence of a magnetic field in case
$p \ne q$ and $c \ne 1.0$.

One of the main issues was the experimental realization of random
fields. Fishman and Aharony \cite{fishaha} showed that the
randomly quenched exchange interactions Ising antiferromagnet in a
uniform field $H$ is equivalent to a ferromagnet in a random field
with the strength of the random field linearly proportional to the
induced magnetization. This identification gave new impetus to the
study of the RFIM, the investigation gained further interest and
was intensified resulting in a large number of publications
(theoretical, numerical, Monte Carlo simulations and experimental)
in the last thirty years. Although much effort had been invested
towards this direction, the only well-established conclusion drawn
was the existence of a phase transition for $d \geq 3$ (d space
dimension), that is, the critical lower dimension $d_{l}$ is 2
after a long controversial discussion \cite{imbrie}, while
many other issues are still unanswered; among them is the order of
the phase transition (first or second order), the universality
class and the dependence of these points on the form of the random
field PDF.The study of RFIM has also highlighted another feature
of the model, that of tricriticality and its dependence on the
assumed distribution function of the random fields. According to
the mean field approximation, the choice of the random field distribution
can lead to a continuous FM/PM boundary as in the single Gaussian
probability distribution, whereas for the bimodal one this
boundary is divided into two parts, an SOPT branch for high
temperatures and an FOPT branch for low temperatures separated by
a TCP at $kT^{t}_{c}/(zJ)=2/3$ and $h^{t}_{c}/(zJ)=(kT^{t}_{c}/(zJ)) \times
\arg\tanh(1/\sqrt{3})\simeq 0.439$ \cite{aharony}, where $z$ is the
coordination number and $k$ the Boltzmann constant, such that for
$T<T^{t}_{c}$ and $h>h^{t}_{c}$ the transition to the FM phase is
first order for the symmetric case $p = \frac{1}{2}$. However, this
behavior is not fully elucidated since in the case of the three-dimensional
RFIM, the high temperature series expansions by Gofman et al
\cite{gofman} yielded only continuous transitions for both
probability distributions, whereas according to Houghton et al
\cite{houghton} both distributions predicted the existence of a
tricritical point, with $h^{t}_{c} = 0.28 \pm 0.01$ and $T^{t}_{c}
= 0.49\pm 0.03$ for the bimodal and $\sigma^{t}_{c} = 0.36 \pm
0.01$ and $T^{t}_{c} = 0.36\pm 0.04$ for the single Gaussian. In
the Monte Carlo studies for $d = 3$, Machta et al \cite{machta},
using single Gaussian distribution, could not reach a definite
conclusion concerning the nature of the transition, since for some
realizations of randomness the magnetization histogram was
two-peaked (implying an SOPT) whereas for other ones three-peaked,
implying an FOPT.

\vspace{-1mm}

In this investigation, we study the EFT applied to the random field
spin-1/2 Ising model on a square lattice with random bond nearest
neighbor interactions, calculating the relevant thermodynamic quantities,
such as critical temperatures, susceptibility, magnetization profiles
with respect to the temperature $T$ and random field $h_{1}$. The paper
is organized as follows: In the next section, the EFT formalism is introduced and
applied to the Ising model in the presence of a random field and random
bonds interactions; the equation of state for the magnetization is derived.
In section $3$, the phase diagrams and magnetization profiles are calculated
as functions of system's parameters; we close with the conclusions in section $4$.

\vspace{-8mm}

\section{Model and formalism}

\vspace{-5mm}

Considering the Hamiltonian (\ref{rham}), each spin is under
the influence of a molecular field $\widetilde{h}_{i}$

\begin{equation}
\widetilde{h}_{i} = \sum_{j=1}^{z} J_{ij}S_{j} + h_{i}
 \label{molfield}
\end{equation}

where $z$ is the coordination number. For the current model the
starting point is the Callen exact spin correlation function
identity \cite{callen}

\begin{equation}
\langle \langle S_{i}\rangle \rangle_{r} = \langle \langle \tanh
\left[ \beta \sum_{j=1}^{z} J_{ij}S_{j} +
\beta h_{i} \right ] \rangle \rangle_{r}          \label{spin1}
\end{equation}

where the summation takes all the nearest neighboring spin sites
of $i$ and $<<\cdot\cdot\cdot>>_{r}$ indicates the thermal
and random configurational averages. Following Honmura and Kaneyoshi
\cite{honkan}, introducing the differential operator $D \equiv
\frac{\partial}{\partial x}$ into (\ref{spin1}) and using also
the van der Waerden identity

\begin{equation}
  e^{J_{ij}S_{j}D} =  \cosh \left (J_{ij}D\right) + S_{j}  \sinh
  \left(J_{ij}D\right)   \label{spin2}
\end{equation}

Eq. (\ref{spin1}) transforms into

\begin{eqnarray}
\langle \langle S_{i}\rangle \rangle_{r} & = & \langle \langle \exp
\left[ \beta D \sum_{j=1}^{z}
J_{ij} S_{j} \right ]\rangle \rangle_{r} \langle\tanh \left [ x + \beta h_{i}\right ]
\rangle_{x=0}   \nonumber  \\
  & = & \langle \langle \prod_{j=1}^{z} \left[\cosh(\beta J_{ij}D) +
  S_{j} \sinh(\beta J_{ij}D) \right]\rangle \rangle_{r} f(x,\beta
  h_{i})_{x=0}  \label{spin3}
\end{eqnarray}

where

\begin{equation}
  f(x,\beta h_{i}) = p \tanh (x+\beta h_{1}) + q  \tanh (x-\beta h_{2})
  \label{spin4}
\end{equation}

Before proceeding to the calculations, we shall consider the
following approximations to make the problem mathematically
tractable, $(a)$ the configurational average of spins and
exchange integral are taken independently,
$ \langle \langle  S_{j}f(J_{ij}) \rangle \rangle \simeq
\langle \langle S_{j} \rangle \rangle \langle f(J_{ij}) \rangle$,
(b) the exchange integrals $J_{ij}$ for different $j\;'s$
are also independent of each other, $ \langle f_{1}(J_{ij})
f_{2}(J_{ik})\rangle = \langle f_{1}(J_{ij}) \rangle \langle
f_{2}(J_{ik}) \rangle $. Under these assumptions the relation
(\ref{spin3}) is written as

\begin{equation}
  \langle \langle S_{i}\rangle \rangle_{r} \simeq \sum _{n=1}^{z} Q_{nz}
  \sum _{j_{1},j_{2} \cdot\cdot\cdot j_{n}=1}^{z} \langle \langle
  S_{j_{1}}S_{j_{2}} \cdot\cdot\cdot S_{j_{n}}\rangle \rangle_{r}
  \label{spin5}
\end{equation}

where all the spins $S_{j_{1}},S_{j_{2}}, \cdot\cdot\cdot, S_{j_{n}}$
are nearest neighbors of $S_{i}$ and the coefficient $Q_{nz}$ is

\begin{equation}
  Q_{nz} = \langle \cosh(\beta D J_{ij})\rangle ^{z-n}
  \langle \sinh(\beta D J_{ij})\rangle ^{n} f(x,\beta h_{i})_{x=0}
  \label{spin6a}
\end{equation}

As far as the $\tanh(x)$ is an odd function of its argument, only
odd $n$ appear in $Q_{nz}$ for the square ($z=4$) lattice.
Considering the PDF (\ref{bimodalJ}) for $J_{ij}$, the averages in
(\ref{spin6a}) are

\begin{eqnarray}
\langle \cosh(\beta D J_{ij})\rangle & = & \cosh(\beta J D)
   \cosh(\beta \Delta D)          \nonumber  \\
\langle \sinh(\beta D J_{ij})\rangle & = &  \sinh(\beta J D)
\cosh(\beta \Delta D)      \label{mean}
\end{eqnarray}

thus (\ref{spin6a}) becomes

\begin{equation}
  Q_{nz} = \cosh^{z-n}(\beta J D) \sinh^{n}(\beta J D)
  \cosh^{z}(\beta \Delta D) f(x,\beta h_{i})_{x=0}   \label{spin6}
\end{equation}

which, for even $z$, takes the form

\vspace{-5mm}

\begin{eqnarray}
Q_{nz} & = & \sum_{\ell =1}^{z/2} b_{\ell}^{(n)} \sinh(2 \ell
\beta J D)   \sum _{\nu =0}^{z/2} |a_{\nu}^{(z)}| \cosh (2\nu
\beta \Delta D) f(x,\beta h_{i})_{x=0}    \nonumber  \\
 & = & \sum_{\ell =1}^{z/2} b_{\ell}^{(n)} \sum _{\nu =0}^{z/2} |a_{\nu}^{(z)}|
 \sinh(2\ell \beta J D)  \cosh (2\nu \beta \Delta D) f(x,\beta h_{i})_{x=0}     \label{spin7}
\end{eqnarray}

\noindent using the operator relation $e^{wD}f(x) = f(x+w)$, we
obtain for the summand in (\ref{spin7})

\vspace{-5mm}

\begin{eqnarray}
\sinh(2\ell \beta J D)  \cosh (2\nu \beta \Delta D) f(x,\beta
h_{i})_{x=0}  =   \nonumber  \\
\frac{1}{4} \mbox{\Large \{} p \,  \mbox{\Large (} \tanh  [ \beta
( 2\ell J + 2\nu \Delta +h_{1})]+  \tanh [ \beta ( 2\ell J - 2\nu \Delta +h_{1})] +  \nonumber  \\
 \tanh [ \beta (2\ell J - 2\nu \Delta -h_{1})] +
 \tanh [ \beta ( 2\ell J + 2\nu \Delta -h_{1})] \mbox{\Large )} +  \nonumber  \\
 q \, \mbox{\Large (} \tanh [ \beta ( 2\ell J + 2\nu \Delta -h_{2})]
 \tanh [ \beta ( 2\ell J - 2\nu \Delta -h_{2})] +  \nonumber  \\
 \tanh [ \beta ( 2\ell J - 2\nu \Delta +h_{2})] +
 \tanh [ \beta ( 2\ell J + 2\nu \Delta +h_{2})]\mbox{\Large )} \mbox{\Large \}}   \nonumber  \\
   \equiv  \frac{1}{4}g_{\ell \nu} (\beta,J,\Delta,h_{1},h_{2})  \label{spin8}
\end{eqnarray}

thus Eq. (\ref{spin7}) is written as

\begin{equation}
  Q_{nz} = \frac{1}{4} \sum_{\ell =1}^{z/2} b_{\ell}^{(n)}
  \sum _{\nu =0}^{z/2}|a_{\nu}^{(z)}| g_{\ell \nu} (\beta,J,\Delta,h_{1},h_{2})  \label{spin9}
\end{equation}

with

\vspace{-5mm}

\begin{eqnarray}
a_{\nu}^{(z)} & = & \frac{2-\delta _{\nu,0}}{2^{z}}
(-1)^{\frac{z}{2}-\nu} \left(%
\begin{array}{c}
  z \\
  \frac{z}{2}-\nu \\
\end{array}%
\right)     \nonumber  \\
 b_{\ell}^{(1)} & = & \frac{2\ell}{z} |a_{\ell}^{(z)}| \nonumber  \\
 b_{\ell}^{(3)} & = & \frac{2\ell}{\prod _{i=0}^{2}(z-i)} [(2\ell)^{2}-(3z-2)]|a_{\ell}^{(z)}| \nonumber  \\
 b_{\ell}^{(5)} & = & \frac{2\ell}{\prod _{i=0}^{4}(z-i)} [(2\ell)^{4}-10(z-2)(2\ell)^{2}+
 15z^{2}-50z+24]|a_{\ell}^{(z)}|     \label{spin10}
\end{eqnarray}

\vspace{-5mm}

Applying this procedure to the linear chain ($z=2$) without
both types of randomness, namely, $h_{1}=0, \Delta=0$, (\ref{spin5})
implies

\begin{equation}
M  = Q_{12} (\langle S_{1}\rangle + \langle S_{2}\rangle) =
2Q_{12}\langle S_{1}\rangle = 2Q_{12} M    \label{eqmagnz2}
\end{equation}

where $Q_{12}= \frac{1}{2} \tanh(2 \beta J)$ from relations
(\ref{spin7}), (\ref{spin8}), (\ref{spin9}), (\ref{spin10}), so
that for the respective critical point we get $2Q_{12}^{c}=1$ or
$\tanh(2 \beta_{c} J)=1$ which implies that

\begin{equation}
(2 \beta_{c} J)^{-1}  = \frac{k_{B}T_{c}}{2J} = 0 \label{crittz2}
\end{equation}

in accordance with the known exact result for the one-dimensional
IM, thus improving on the MFA result that is $(k_{B}T_{c})/J = 2$.

For the disordered square lattice ($z=4$), the respective $Q_{nz}$ functions are

\begin{eqnarray}
 Q_{14} & = &  \frac{1}{2^{8}} ( 2G_{1} + G_{2} )  \nonumber  \\
 Q_{34} & = &  \frac{1}{2^{8}} (-2G_{1} + G_{2} )    \label{Kz4}
\end{eqnarray}

where $ G_{k} = 3g_{k0}+4g_{k1}+g_{k2} $ and the g's functions are defined
in (\ref{spin8}). The resulting equation for the equilibrium magnetization
from (\ref{spin5}) is

\begin{equation}
m = 4Q_{14}m + 4Q_{34}m^{3}     \label{magnz4}
\end{equation}

which admits two solutions, the paramagnetic one, $m=0$, and the ferromagnetic one
given by

\begin{equation}
   m = \pm \sqrt{\frac{1 - 4Q_{14}} {4Q_{34}} }    \label{magnz4sol}
\end{equation}

whereas the critical boundary characterizing the ferromagnetic/paramagnetic
phases is determined by the condition $m=0$ and results as a solution to the
equation $4Q_{14} = 1$, which, in case both the random field strength $h_{1}$
and exchange integral deviation $\Delta$ vanish, then it converts into

\begin{equation}
  2 \tanh(2\beta_{c}J) + \tanh(4\beta_{c}J)=2     \label{crittempd2}
\end{equation}

as was also found in \cite{satsa}, and the critical temperature results as
$k_{B}t_{c}/J\cong 3.0898\ldots$, which is closer to the exact one
$k_{B}t_{c}^{exact}/J= 2/\ln(1+\sqrt{2})$ than the MFA one $k_{B}t_{c}^{MFA}/J=4$.
At this point, it has to be noted that the decoupling
scheme recalled in Eq. (\ref{magnz4}) behaves as a better approximation than
the one in the MFA, since within the EFT framework the kinematics relations
are treated exactly ($\sigma^{2}=1$) through the van der Waerden identity;
in the present case, EFT neglects correlations only between different spin variables,
whereas the MFA neglects any kind of correlation, namely, the self- and multi-spin ones.

\vspace{-6mm}
\section{Numerical results. Phase diagrams. Magnetization profiles}
\vspace{-5mm}

\begin{figure}[htbp]
\begin{center}
\includegraphics*[height=0.25\textheight]{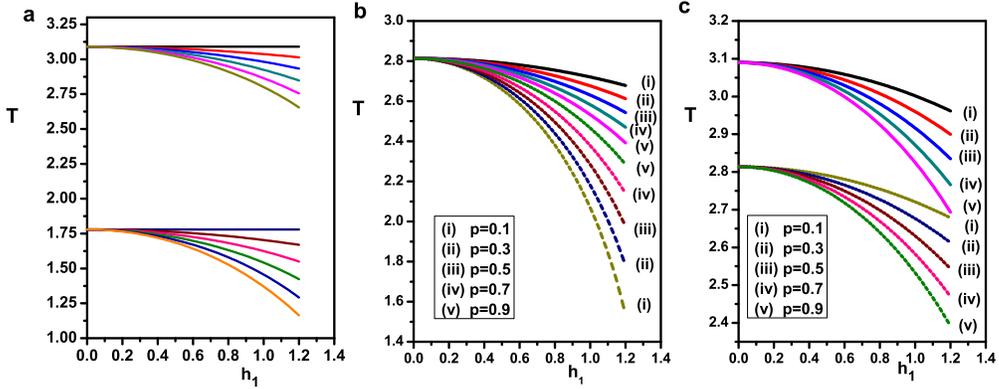}
\caption{\label{figa} (Color online) Random field $h_{1}$ variation
of the critical temperature $T$ - phase diagram ($h_{1}-T$).
In panel (a), for the upper group of graphs $c=d=0.0$,
for the lower group $c=0.0\; d=1.0$.
In each group from up to down $p=0.0, 0.2, 0.4, 0.6, 0.8, 1.0$.
In panel (b), d is fixed as $d=0.5$, $c=0.5$ (upper solid graphs),
and $c=1.5$ (lower dashed graphs), labelled by the respective $p-$values.
The solid curves are arranged in increasing $p$ from up to down,
whereas the dashed ones are arranged in increasing $p$ from down to up.
In panel (c), c is fixed as $c=0.5$, $d=0.00001$ upper solid graphs,
and $d=0.5$ lower dashed graphs. The plots in both groups are labelled
by the respective $p-$values. $T$ in units of $J/k$.}
\end{center}
\end{figure}

An important feature of the magnetic or fluid systems is their phase diagram,
the temperature against a suitably chosen parameter, which, in the present
case, allows the investigation more clearly of the effects of randomness
itself on the transition temperature of the random system. In the current study,
there are various parameters, such as $c,\Delta,p,h_{1}$, so that
we can introduce a variety of phase diagrams by plotting the
temperature with respect to one of these parameters keeping the
other ones fixed. However, in the numerical calculations to follow
the main quantities $\beta, \Delta, h_{1}, h_{2}$ are written with
respect to the average value $J$ of the exchange integral, that is,
$\beta \equiv J \beta, d \equiv \Delta/J, h_{1} \equiv h_{1}/J$,
$h_{2} \equiv h_{2}/J$, scaled quantities; this choice influences
Eqs. (\ref{bimodalJ}) and (\ref{bimodal}): the PDF in (\ref{bimodalJ})
is written as $P(J_{ij})=\frac{1}{2}\{\delta(J_{ij}-(1+d)) + \delta (J_{ij}-(1-d))\}$
so that by choosing $d = 1.0$ the former PDF changes into
$P(J_{ij})=\frac{1}{2}\{\delta(J_{ij}-2) + \delta (J_{ij})\}$ implying that
some of the bonds are missing, diluted bonds. If we also set $c=0.0$ in
Eq. (\ref{bimodal}), then some of the sites are occupied either by non magnetic
particles or are empty, since the respective PDF (\ref{bimodal}) converts into
$P(h_{i})=p\delta(h_{i}-h_{1}) + q \delta (h_{i})$. These cases are shown in
Fig~\ref{figa}(a) with each graph labelled by the respective site probability $p$
with the graphs forming two groups: the upper one corresponding to $c=d=0.0$
(site diluted system with vanishing exchange integral deviation) and the lower
group corresponding to $c=0.0, d=1.0$, site and bond diluted system.
The choice $d=1.0$ implies that the deviation
is of the order of the exchange integral ($d\sim|J|$) so that some bonds may
vanish according to the Eq. (\ref{bimodalJ}) whereas the remaining ones are
strengthened. Also, other phase diagrams are presented in Figs.~\ref{figa}(b,c)
with each line labelled by a p-value.
In Fig.~\ref{figa}(b), with d fixed as $d=0.5$, the five upper solid lines correspond
to $c=0.5$, whereas the five lower dashed ones to $c=1.5$. The topmost solid curve $(i)$
is for $p=0.1$ and as $p$ increases towards $p=1.0$, (curve $(v)$) the
respective curve lies below that corresponding to a lower value for $p$,
the higher the p-value the lower the respective curve lies; this causes a
gradual reduction of the ferromagnetic region. As far as the second group is
concerned in Fig.~\ref{figa}(b) (dashed ones),
an inversion of the order of the curves occurs in comparison to the upper solid ones with
respect to $p$, that is, now the lowermost one $(i)$ corresponds to $p=0.1$,
whereas as $p$ increases towards to $p=1.0$ $(v)$, the respective line lies
above that with a lower value for $p$, so that, the higher the $p$-value
the higher the curve lies, thus an enlargement of the ferromagnetic region takes place.
However, before the inversion of the order of
the lines takes place in panel (b), one observes coincidence of the phase-diagram
boundaries for the all the $p$-values for $d=0.5, c=1.0$; in this case $h_{1}=|h_{2}|$.
However, the order of the curves is unaltered if the temperature $T$ is
drawn with respect to the random field $h_{1}$ for a fixed value of c ($c=0.5$),
whereas $d$ varies as $d=0.00001$ (solid curves) and $d=0.5$ (dashed ones),
Fig.~\ref{figa}(c), with the graphs for the same $d-$value forming a separate group,
whereas in Fig.~\ref{figa}(b) they form a single group.
A distinctive feature resulting from the plots in Fig.~\ref{figa} is that all
the lines of the phase diagrams coincide for small values of $h_{1}$, beginning
from the same point on the T-axis for $h_{1}=0.0$, implying that the critical
temperature is the same regardless of $c$ and $p$ but depending only on $d$.
Also, in this figure the critical temperature for $h_{1}=0.0$ is greater the
smaller the $d-$value is. All the critical lines, separating the FM and PM phases,
are of second order phase transitions.

\begin{figure}[htbp]
\begin{center}
\includegraphics*[height=0.30\textheight]{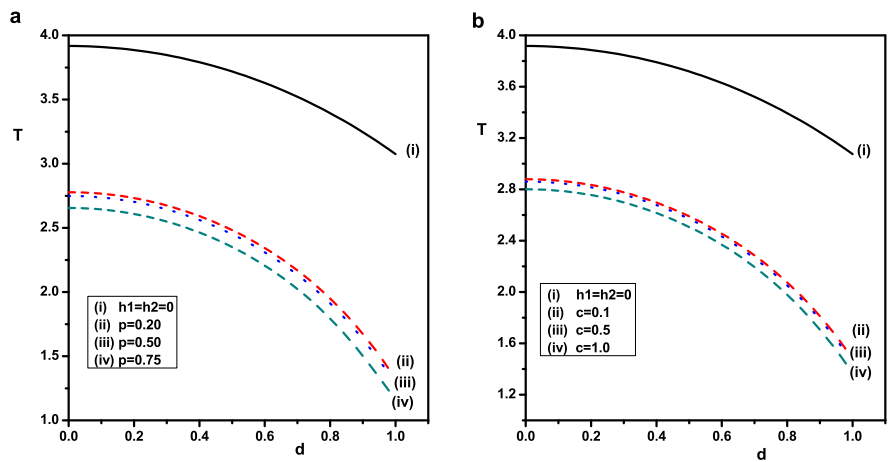}
\caption{\label{figb} (Color online) The critical temperature $T$
as a function of the exchange-integral deviation $d$.
In both panels, the upper solid graph (i) corresponds to $h_{1}=0.0$
irrespectively of $p$ (panel (a)) or $c$ (panel (b).
In panel (a), the lower graphs  are for fixed $c$ and $h_{1}$ as
$c=0.1$, $h_{1}=1.0$ and labelled by the selected $p-$values.
In panel (b), the lower graphs are for fixed $p$ and $h_{1}$ as
$p=0.75$, $h_{1}=1.0$ and labelled by the selected $c-$values.
In both panels the higher the labeling value the smaller
the extent of the ferromagnetic phase.}
\end{center}
\end{figure}

The dependence of the critical temperature $T_{c}$ on the exchange-integral
deviation $d=\Delta/J$ appears in Fig.~\ref{figb} for fixed random field $h_{1}=1.0$
as well as for the case $h_{1}=0.0$ for comparison. An overall feature of both panels
is that each line starts from a different point on the $T-$axis for $d=0.0$ as
compared to Figs.~\ref{figa}. In Fig.~\ref{figb}(a), $c$ was fixed as $c=0.1$ with
the respective curves labelled by a $p-$value, whereas in Fig.~\ref{figb}(b), the
constant quantity was $p$ as $p=0.75$ with the individual curves labelled by a
$c-$value. In both panels the boundary line for $h_{1}=0.0$ lies above the ones
for non zero values of $h_{1}$, thus defining the widest ferromagnetic
phase space in comparison to that for non zero $h_{1}'s$ implying that the effect
of including the random field is to reduce the ferromagnetic region; also the
boundaries are arranged one below the other in ascending order either for $p$ or $c$
so that each time $p$ or $c$ is increased the respective ferromagnetic region is
reduced in comparison to that with a smaller $p$ or $c$.

\begin{table}
\caption{\label{table1} The critical field $h_{1,c}^{(1)}-$ values for
constant values $c=0.5, d=0$ (second column) and $h_{1,c}^{(2)}-$ values
for $c=d=0.5$ (third column) with respect the parameter $p$. $h_{1,c}^{(3)}$
refers to the parameter $d$ as a variable with constant values $c=0.5, p=0.1$
(fifth column), whereas $h_{1,c}^{(4)}$ refers to the parameter $c$ as a variable
with constant values $d=0.5, p=0.1$  (seventh column).}
\begin{tabular}{ccccccccc}
\hline $p$ & $h_{1,c}^{(1)}$ &  $h_{1,c}^{(2)}$ & $d$& $h_{1,c}^{(3)}$ &$c$&  $h_{1,c}^{(4)}$ \\
\hline
$0.0$ & $6.736241$  & $6.194181$ & $0.0$  & 5.904740 & $0.0$  & 11.124309 &  \\
$0.1$ & $5.900871$  & $5.406162$ & $0.1$  & 5.888921 & $0.1$  &  9.678039 &  \\
$0.2$ & $4.421102$  & $4.867123$ & $0.2$  & 5.840427 & $0.2$  &  8.624253 &  \\
$0.3$ & $4.849948$  & $4.435107$ & $0.3$  & 5.758102 & $0.3$  &  7.543436 &  \\
$0.4$ & $4.488253$  & $4.099344$ & $0.4$  & 5.640081 & $0.4$  &  6.365760 &  \\
$0.5$ & $4.190953$  & $3.816993$ & $0.5$  & 5.482039 & $0.5$  &  5.482039 &  \\
$0.6$ & $3.940628$  & $3.572388$ & $0.6$  & 5.277713 & $0.6$  &  4.768053 &  \\
$0.7$ & $3.725971$  & $3.363212$ & $0.7$  & 5.017263 & $0.7$  &  4.189970 &  \\
$0.8$ & $3.538211$  & $3.190681$ & $0.8$  & 4.686686 & $0.8$  &  3.716970 &  \\
$0.9$ & $3.372639$  & $3.053922$ & $0.9$  & 4.269706 & $0.9$  &  3.324434 &  \\
$1.0$ & $3.224684$  & $2.894779$ & $1.0$  & 3.768020 & $1.0$  &  2.993567 &  \\
\hline \hline
\end{tabular}
\end{table}

The critical field $h_{1,c}$, the field for which the critical temperature vanishes,
depends on the thermodynamic route followed, since the current random system contains
various parameters, namely, $c, d, p$; consequently, in order to estimate $h_{1,c}$,
two of these parameters are fixed, only the remaining one varies. Initially, $c$ and $d$
are kept fixed as $c=0.5, d=0$ and later $c=d=0.5$; the values of the respective
critical field  $h_{1,c}$ appear in Table~\ref{table1} for selected $p-$values for both
choices. In the former case, these values are fitted by the sixth-degree polynomial
$P(x)=6.73612-9.97119x+19.00728x^{2}-30.76579x^{3}+33.56416x^{4}-20.72121x^{5}+5.37542x^{6}$,
whereas in the latter case by the fifth-degree polynomial
$P(x)=6.19323-9.39129x+18.25539x^{2}-27.81661x^{3}+23.72744x^{4}-8.07256x^{5}$,
with $x\equiv p$ in both cases. Also, in the same Table, there appear the values for
the critical field $h_{1,c}^{(3)}$ (fifth column) corresponding to the constant quantities
$c=0.5, p=0.1$ with parameter $d$ as a variable, for which the best fit polynomial is
$P(x)=5.90487-1.6687x^{2}+0.39594x^{3}-0.8657x^{4}$ ($x\equiv d$);
moreover, in the seventh column there appear the values for the critical field
$h_{1,c}^{(4)}$ corresponding to the constants $d=0.5, p=0.1$ and parameter $c$ as a variable, with best fit  polynomial $P(x)=11.12544-14.35749x+6.410899x^{2}-0.155474x^{4}$  ($x\equiv c$).

\begin{figure}[htbp]
\begin{center}
\includegraphics*[height=0.3\textheight]{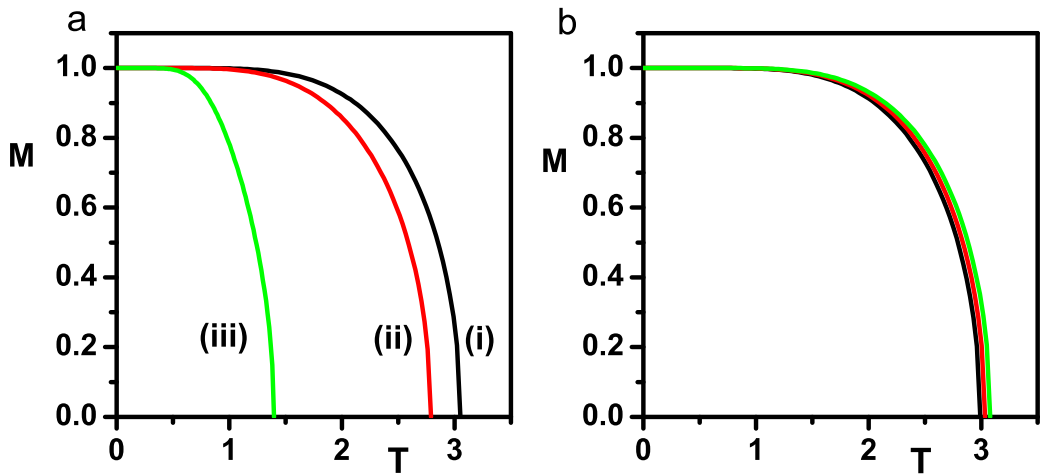}
\caption{\label{figc}(Color online) The thermal dependence of magnetization
$M(T,h_{1})$, order parameter.
In panel $(a)$, the fixed parameters are $d=0.1, p=0.1, h_{1}=1.0$;
each graph is labelled by the random field ratio
$c=h_{2}/h_{1}$: $c=0.1$ graph $(i)$, $c=1.0$ graph  $(ii)$ and
$c=2.0$ graph $(iii)$.
In panel $(b)$, the fixed parameters are $c=3.0, d=0.1$;
each graph is labelled by $p$ and $h_{1}$ as:
innermost graph $p=0.1, h_{1}=0.2$, middle one $p=0.6, h_{1}=0.2$
and outermost $p=0.1, h_{1}=0.0$.}
\end{center}
\end{figure}

In addition to the phase diagram, another important quantity is the magnetization;
its thermal behavior appears in Fig.~\ref{figc} for various values of the
system's parameters $c,d,p,h_{1}$, resulting by solving numerically the Eq. (\ref{magnz4}).
In panel (a) the parameters $d,p,h_{1}$ are fixed as $d=0.1, p=0.1, h_{1}=1.0$;
the individual graphs are labelled by the random field ratio $c=h_{2}/h_{1}$,
the outermost graph ($i$) corresponds to $c=0.1$, the middle one ($ii$) to $c=1.0$
and the innermost one ($iii$) to $c=2.0$. The magnetization exhibits its normal
behavior but it is stronger for the smaller value of this parameter $c=0.1$
(for $c=0.0$ the respective graph is very close to that for $c=0.1$ due
to the finiteness of the calculations), which is in accordance with the respective
plot in Ref. \cite{kantam} although in the latter publication the random field is
not included. In the latter case ($c=2.0$) the stronger negative random field
$h_{2}=ch_{1}$ yields smaller magnetization because this field competes strongly the
first term in the Hamiltonian (\ref{rham}), which favors the parallel orientation
of the spins. In panel (b) the parameters $c,d,h_{1}$ are fixed as $c=3.0, d=0.1, h_{1}=0.2$,
the plots are labelled by the site probability $p=0.1$ (innermost graph) and $p=0.6$
for the middle graph; for comparison the case for the lack of the random field was
also included, which is the outermost graph corresponding to $c=3.0, d=0.1, p=0.1, h_{1}=0.0$. For small values of the temperature, the magnetization values are constant,
independent of the relative parameters. 

A susceptibility-like quantity $\chi_{T}(h_{1})$ can also be calculated as the first order
derivative of the magnetization with respect to the applied external random magnetic field,
namely

\begin{equation}
\chi_{T}(h_{1}) = \frac{\partial m}{\partial h_{1}} = \frac{\partial m}{\partial
h_{0}} \frac{dh_{0}}{dh_{1}} = \frac{1}{J} \frac{\partial m}{\partial h_{0}}  \label{susc1}
\end{equation}

so that

\begin{equation}
   J\chi_{T}(h_{1}) = \frac{\partial m}{\partial h_{0}} =
   \frac{4m\frac{\partial Q_{14}}{\partial h_{0}} +
   4m^{3}\frac{\partial Q_{34}}{\partial h_{0}}}
   {1-4Q_{14}-12m^{2}Q_{34}}      \label{susc2}
\end{equation}

or, for calculational purposes,

\begin{equation}
  \mbox{\Large (} J\chi_{T}(h_{1}) \mbox{\Large )}^{-1}= \frac{1-4Q_{14}-12m^{2}Q_{34}}
  {4m\frac{\partial Q_{14}}{\partial h_{0}} + 4m^{3}\frac{\partial Q_{34}}{\partial h_{0}}}
    \label{susc3}
\end{equation}

since $h_{0} = Jh_{1}$, although the product ($Jh_{1}$) was
earlier set as $h_{1}$, now we divert for calculational purposes.

\begin{figure}[htbp]
\begin{center}
\includegraphics*[height=0.26\textheight]{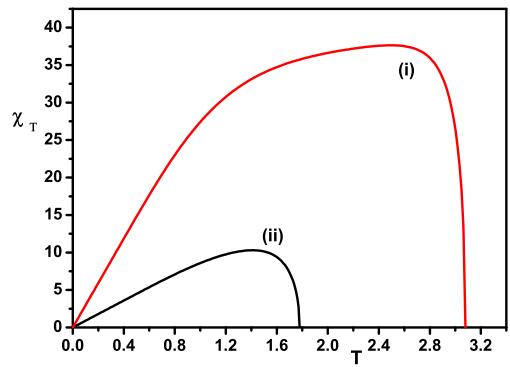}
\caption{\label{figd}(Color online)  The inverse magnetic
susceptibility-like vs temperature $T$, with constant $c=p=h_{1}=0.1$;
graph (i) corresponds to $d=0.1$ and graph (ii) to $d=1.0$.}
\end{center}
\end{figure}

The temperature dependence of the inverse susceptibility-like is shown in
Fig.~\ref{figd} for selected values of the parameters $c,d,p,h_{1}$.
In both graphs, the susceptibility diverges twice, once at the zero
temperature ($T=0$, finite clusters contribution) and once again at
the respective critical point $T_{c}$, infinite cluster contribution.
Although both graphs increase smoothly for small $T-$values, they
decrease steeply for larger ones. A similar behavior for the
susceptibility was also found by Kaneyoshi et al \cite{kanhon}.

\vspace{-6mm}

\section{Conclusions and discussions}

\vspace{-6mm}

In the current investigation we have determined the phase diagrams,
critical temperatures, magnetization profiles and susceptibility of
the nearest-neighbor spin-1/2 Ising model on a square lattice, in
the presence of competing random fields as well as fluctuating random
bonds via EFT, based on the differential operator technique
and Callen identity. The topology of the system is taken into account
through the coordination number. The EFT method is an improvement on
the MFA, since it provides vanishing critical temperature for
one-dimensional systems, also for two-dimensional systems the
respective critical temperature is closer to the exact one than that
of the MFA. A variety of phase diagrams were obtained with respect to
the various system's parameters, namely, $c,d,h_{1},p$. The extent
of the ferromagnetic region does not show systematic variation on
varying the system's parameters; it varies non monotonically for
constant $d$ and varying $c$, whereas in other circumstances it varies
monotonically. The magnetization profile was determined with respect to
temperature as well as the random field $h_{1}$.  An immediate
result of the presence of the random field is to reduce the extent
of the ferromagnetic region, in general, as well as numerical value of the
magnetization as $c$ increases because of the strengthening of the random
field $h_{2}=ch_{1}$ and the subsequent increased tendency of this field to
prevent spins to attain the parallel configuration and the competition with
the other random field $h_{1}$.

Although the EFT technique improves on the MFA, it presents the same
shortcomings in the critical region (non-classical region) concerning
the critical exponents like the MFA and Landau theory, because the
fluctuations occurring in the critical region, as the transition
temperature is approached, become important and the non-classical
behavior is observed; these fluctuations are not taken under
consideration properly by EFT or MFA, thus a relative criterion,
called Ginzburg criterion, determines how close to the transition
temperature the true critical behavior is revealed \cite{ginzburg}.
This criterion relies on any thermodynamic
quantity but the specific heat is usually considered for
determining the critical region around $T_{c}$ where the mean
field solution cannot describe correctly the phase transition. The
MFA is valid for lattice dimensionality greater than the
upper critical dimension $d_{u}=4$ in case of presence of only
thermal fluctuations. However, in the current case the presence of
random fields enhances fluctuations causing the critical region to
be wider than the one due only to the thermal fluctuations
 and the upper critical dimension is increased by $2$ to $d_{u}=6$
 \cite{kaufmankardar,nielsen}. Occasionally the non classical
region is extremely narrow so that the respective critical
behavior expected by Landau or MFA is observed because the fluctuation
region is very narrow and hardly accessible for experimental
observation; such a system is the weak-coupling superconductor in
three dimensions for which the respective non classical region is
$|t_{CR}|\leq 10^{-16}$ ($t_{CR}$ is the reduced temperature).
However, on reducing the space dimension as in the case of the
weak-coupling superconductor in two dimensions, the critical
exponents have their classical values up to $|t_{CR}|= 10^{-5}$,
thus the reduction of the space dimensionality causes serious
repercussions on the behavior of the physical system; on the
contrary, for the superfluid helium transition the classical
region extends up to $|t_{CR}|\leq 1.0$ so that fluctuations are
detectable \cite{patapok,ivan,domb,goldenfeld}. In addition to
superconductivity, the extent of the non classical region for the
ferroelectric system triglycine sulfate (TGS) is relatively small
and its critical exponents have the respective classical values up
to $|t_{CR}|= 1.5\times 10^{-5}$ \cite{fe1,fe2,fe3}.

The results obtained in the current investigation by using the EFT
provide a basis for a comprehensive analysis by more sophisticated 
methods. However, they are of no less importance, since they show, 
nevertheless, the expected phenomena to be observed.

\vspace{-9mm}

\ack{This research was supported by the Special Account for
Research Grants of the University of Athens ($E\Lambda KE$) under
Grant No. 70/4/4096.}

\end{document}